\documentclass[11pt, draftclsnofoot, onecolumn]{IEEEtran}

\usepackage{amsmath}
\usepackage{booktabs} 
\usepackage{multirow}
\usepackage{array}
\usepackage{color}

\usepackage{cite}

\usepackage{mathtools}
\usepackage{cite}
\usepackage[justification=centering]{caption}
\usepackage{graphicx}
\usepackage{textcomp}
\usepackage{xcolor}
\usepackage{booktabs}

\usepackage{epstopdf}
\usepackage{amsmath,amsthm, amssymb}
\usepackage{threeparttable}
\usepackage{bm}
\usepackage{multirow}
\usepackage{booktabs}
\usepackage{color}
\usepackage{tabularx}
\usepackage{setspace}
\usepackage{verbatim}
\usepackage{stfloats}
\usepackage{caption}
\usepackage{float}
\usepackage{url}

\usepackage{subfigure}

\usepackage{setspace}

\usepackage{epstopdf}

\usepackage{amsmath}
\usepackage{stfloats}
\usepackage{amsmath}
\allowdisplaybreaks[0]

\makeatletter
\newif\if@restonecol
\makeatother

\usepackage[linesnumbered,ruled,vlined]{algorithm2e}
\usepackage{algpseudocode}
\usepackage{amsmath}

\usepackage{fancyhdr}

\usepackage{hyperref}
\makeatletter
\def\makeLineNumberLeft{%
  \linenumberfont\llap{\linenumbersep\LineNumber\hss}}
\def\linenumber{%
  \@tempcnta\c@linenumber
  \@whilesw\ifnum\@tempcnta>\tw@\do{%
    \advance\@tempcnta\m@ne
    \LineNumber\hss}%
  \ifnum\c@linenumber<\m@ne
  \else
    \makeLineNumberLeft
  \fi}
\makeatother

\begin{document}
\pagestyle{fancy}
\fancyhf{}
\fancyhead[L]{IEEE XXX, VOL. XX, NO. XX, XXX 2023}

\captionsetup{font={small}}
\title{Delay-sensitive Task Offloading in Vehicular Fog Computing-Assisted Platoons}

\author{Qiong Wu,~\IEEEmembership{Senior Member,~IEEE}, Siyuan Wang, Hongmei Ge \\Pingyi Fan, ~\IEEEmembership{Senior Member,~IEEE}, Qiang Fan, and Khaled B. Letaief,~\IEEEmembership{Fellow,~IEEE}
\thanks{This work was supported in part by the National Natural Science Foundation of China under Grant No. 61701197, in part by the open research fund of State Key Laboratory of Integrated Services Networks under Grant No. ISN23-11, in part by the National Key Research and Development Program of China
under Grant No. 2021YFA1000500(4), in part by the 111 Project under Grant No. B23008. (Corresponding Author: Hongmei Ge)

Qiong Wu, Siyuan Wang and Hongmei Ge are with the School of Internet of Things Engineering, Jiangnan University, Wuxi 214122, China, and also with the State Key Laboratory of Integrated Services Networks (Xidian University),  Xi'an 710071, China (e-mail: qiongwu@jiangnan.edu.cn, \{siyuanwang, hongmeige\}@stu.jiangnan.edu.cn)

Pingyi Fan is with the Department of Electronic Engineering, Beijing National Research Center for Information Science and Technology, Tsinghua University, Beijing 100084, China (e-mail: fpy@tsinghua.edu.cn).

Qiang Fan is with Qualcomm, San Jose, CA 95110, USA (e-mail: qf9898@gmail.com)

K. B. Letaief is with the Department of Electrical and Computer Engineering, the Hong Kong University of Science and Technology (HKUST), Hong Kong, and also with the Pengcheng Laboratory, Shenzhen 518055, China (e-mail:eekhaled@ust.hk).

}
}


\maketitle

\begin{abstract}
Vehicles in platoons need to process many tasks to support various real-time vehicular applications. \textcolor{black}{When a task arrives at a vehicle}, the vehicle may not process the task due to its limited computation resource. In this case, it usually requests to offload the task to other vehicles in the platoon for processing. \textcolor{black}{However, when the computation resources of all the vehicles in the platoon are insufficient, the task cannot be processed in time through offloading to the other vehicles in the platoon.} \textcolor{black}{Vehicular fog computing (VFC)-assisted platoon} can solve this problem through offloading the task to the VFC which is formed by the vehicles driving near the platoon. Offloading delay is an important performance metric, which is impacted by both the offloading strategy for deciding where the task is offloaded and the number of the allocated vehicles in VFC to process the task. Thus, it is critical to propose an offloading strategy to minimize the offloading delay. \textcolor{black}{In the VFC-assisted platoon system}, vehicles usually adopt the IEEE 802.11p distributed coordination function (DCF) mechanism while having various computation resources. Moreover, when vehicles arrive and depart the VFC randomly, their tasks also arrive at and depart the system randomly. In this paper, we propose a \textcolor{black}{semi-Markov} decision process (SMDP) based offloading strategy while considering these factors to obtain the maximal long-term reward reflecting the offloading delay. \textcolor{black}{Our research provides a robust strategy for task offloading in VFC systems, its effectiveness is demonstrated through simulation experiments and comparison with benchmark strategies.}

\end{abstract}

\begin{IEEEkeywords}
Platoons, vehicular fog computing, offloading, delay.
\end{IEEEkeywords}

\IEEEpeerreviewmaketitle

\section{Introduction}
\IEEEPARstart{W}{ith} the development of autonomous driving technology, platoons have been widely studied to improve the road safety. A platoon is composed of several autonomous driving vehicles such as a leader vehicle and other member vehicles. The leader vehicle controls the velocity, acceleration and driving direction of the platoon. The member vehicles along the same lane follow the leader vehicle one by one \cite{r1,wu13}. The on-board equipments of vehicles in the platoon such as radar and lidar collect a large amount of data to support various real-time vehicular applications such as automatic navigation, multimedia entertainment, high definition (HD) map and precise positioning\cite{r2,wq01,wu11,wu12}. These data are usually redundant and thus vehicles in the platoon could compute and analyze the data to extract useful information. Therefore, vehicles in platoons may have many tasks to deal with.

When a task arrives at a vehicle in a platoon, the vehicle may not be able to process the arrived task due to its limited computation resource. In this case, it usually offloads the task to the other vehicles in the platoon for assistant processing, whose result can be sent back to the vehicle. However, when the computation resources of the platoon are still insufficient, the task cannot be processed in time by offloading to the platoon. \textcolor{black}{We have utilized a vehicular fog computing (VFC)-assisted system to tackle this problem,} where the VFC is formed by the personal vehicles near the platoon. \textcolor{black}{In VFC-assisted platoon}, one vehicle is able to request to offload tasks to the VFC to assist the platoon for processing \cite{r8,TNSM1,TNSM2}. When the requested vehicle offloads a task to the VFC, it transmits the task to the leader vehicle. Afterwards, the system determines how many vehicles in the VFC to process and then the leader vehicle divides the task into the corresponding number of subtasks. Then the leader vehicle transmits the subtasks to the corresponding vehicles in the VFC one by one. After that the vehicles process the assigned subtasks and send back the computing result to the leader vehicle who further forwards the computing result to the requested vehicle.

Offloading delay is an important metric for the autonomous vehicles \textcolor{black}{in the VFC-assisted platoon system} \cite{r9}. If the offloading delay is large, the requested vehicle in the platoon would take a long time to receive the result, resulting in unsatisfying the requirements of vehicular applications. \textcolor{black}{The offloading delay includes transmitting delay, computing delay, and backhaul delay.} In fact, transmitting delay represents the time that the requested vehicle transmits the task to other vehicles for assistant processing; computing delay represents the time that the vehicles process the task and the backhaul delay is the time that vehicles send back the computing result.
The offloading delay is impacted by both the locations where tasks are offloaded and the number of vehicles allocated for task processing. On the one hand, determining the place where the task is offloaded will affect the offloading delay. Specifically, offloading the task to the VFC will incur a higher transmitting delay, while the VFC can provide more computation resources to process the task, which would cause a lower computing delay.
Offloading the task to the platoon will incur a lower transmitting delay, \textcolor{black}{while the computation resources in the platoon are limited}, which would cause a higher computing delay.
Hence, it is important to determine where the task is offloaded to minimize the offloading delay. On the other hand, the number of the allocated vehicles in VFC to process the task will also affect the offloading delay. Specifically, if more vehicles are chosen for offloading, the sufficient computation resources reduce the computing delay, while more subtasks would be divided and transmitted to the vehicles in the VFC in turn, which increases the transmitting delay. Hence, it is another important problem to determine how many vehicles in the VFC to process tasks to minimize the offloading delay. As far as we know, there is no work to design a task offloading strategy \textcolor{black}{in the VFC-assisted platoon}, which motivates us to conduct this work.

\textcolor{black}{In the VFC-assisted platoon system}, vehicles usually adopt the IEEE 802.11p distributed coordination function (DCF) mechanism to offload tasks \cite{r4,wq02,r5, r6, r7}. The computation resources of vehicles in the platoon are usually different. Meanwhile, tasks arrive and depart the system randomly, and the vehicles also arrive and depart the VFC randomly. These factors pose challenges to design the optimal offloading strategy to obtain the maximal the long-term reward related to offloading delay of the system. This paper jointly considers the 802.11p DCF mechanism and some factors, i.e., heterogeneous computation resource of vehicles in the platoon, the random task arrival and departure as well as the random arrival and departure of vehicles in the VFC,
to design an \textcolor{black}{semi-Markov} decision process (SMDP) based offloading strategy by maximizing the long-term reward \textcolor{black}{in the VFC-assisted platoon}. The main contributions of this paper are summarized as follows.

\begin{itemize}
\item[1)] \textcolor{black}{We propose an offloading strategy for a VFC-assisted platoon system. This strategy takes into account several factors, i.e., the heterogeneous computation resources of vehicles in the platoon, the random arrival and departure of task, the random arrival and departure of the vehicles in the VFC, and the 802.11p DCF mechanism.}
\item[2)] We adopt the SMDP to model the task offloading process and design the SMDP framework including state, action, reward and transition probability, where the transmitting delay is derived according to the 802.11p DCF mechanism to determine the reward.
\item[3)] We propose an iterative algorithm to solve the SMDP model to obtain the optimal offloading strategy. \textcolor{black}{Extensive experiments demonstrated that the proposed offloading strategy outperforms the baseline strategy.}
\end{itemize}

The rest of this paper is organized as follows. Section II reviews the related work. Section III describes the system model briefly. Section IV sets up the SMDP framework to model the offloading process \textcolor{black}{in VFC-assisted platoon}. Section V solves the SMDP model by an iterative algorithm. Section VI presents the simulation results. Section VII concludes this paper.

\section{Related Work}
\textcolor{black}{In this section, we review the state-of-art works on the offloading in platoons and offloading in VFC.}

\subsection{Task offloading in platoons}
In recent years, some works designed the task offloading strategy in platoon to optimize various performances.
In \cite{w1}, Fan \textit{et al.} designed the task offloading strategy in the mobile edge computing (MEC)-assisted platoon. They adopted the directed acyclic graph to model the task offloading process, and then employed the \textcolor{black}{Lagrangian} relaxation-based aggregated cost (LARAC) algorithm to minimize the cost of the task offloading.
In \cite{w2}, Ma \textit{et al.} considered the change of vehicle velocity in the platoon and proposed a task offloading and task take-back strategy. The reinforcement learning is adopted to enhance the offloading efficiency and avoid the link disconnection resulting from task processing failures.
In \cite{w3}, Hu \textit{et al.} designed the task offloading strategy in the MEC -assisted platoon under the stable condition of task queues. The Lyapunov optimization algorithm is employed to reduce the energy consumption of task execution and improve the offloading efficiency considering transmitting delay as well as computing delay.
In \cite{w4}, Du \textit{et al.} considered the sensing, computation, communication and storage resources of a single autonomous vehicle and proposed a communication strategy for autonomous driving platoon. In addition, they adopted the genetic algorithm to design the offloading strategy to achieve the minimized processing delay.
In \cite{w5}, Zheng \textit{et al.} proposed a platoon task offloading strategy based on the dynamic non-orthogonal multiple access (NOMA). The long-term energy consumption is minimized by optimizing the allocation of both the communication and computation resource based on the Lyapunov and block successive upper bound minimization (BSUM) methods.
In \cite{w23}, Xiao \textit{et al.} considered a scenario where the requested vehicle offloads tasks to the vehicles in the platoon and proposed a resource allocation strategy based on the requested vehicle's service pricing to effectively utilize the resources in the platoon.
\textcolor{black}{However, the computation resource in a platoon is limited. When the resources in the platoon are not enough to process the task, a potential threat will be posed for safety. In this situation, the vehicles driving near the platoon may form a VFC to provide computation resource.}

\subsection{Task offloading in VFC}
Many works have studied task offloading in the VFC.
In \cite{w6}, Liao \textit{et al.} considered the uncertain information and designed a  task offloading strategy in the VFC to get the minimized total network delay.
In \cite{w7}, Zhou \textit{et al.} considered a two-stage VFC framework and designed a task offloading strategy according to the price-based matching algorithm to minimize the long-term task offloading delay.
In \cite{w8}, Shi \textit{et al.} proposed a task offloading strategy, where vehicles are likely to share idle computation resources by dynamic pricing strategy to maximize the utility of offloading tasks.
In \cite{w9}, Yadav \textit{et al.} considered the vehicular node mobility and end-to-end latency deadline constraints and designed an energy-efficient dynamic computation offloading and resources allocation strategy (ECOS) in the VFC to minimize energy consumption and service latency.
In \cite{w10}, Xie \textit{et al.} considered the parallel computation task offloading problem in the VFC and proposed an offloading strategy to reduce the service time and improve the amount of the finished tasks by using the hidden \textcolor{black}{Markov model.}
In \cite{w11}, Lqbal \textit{et al.} designed an offloading strategy to improve the performance of queuing delay, end-to-end delay and task completion rate based on the computing capacity and the workload of vehicles in the VFC.
In \cite{w12}, Misra \textit{et al.} proposed a task offloading strategy in a software-defined VFC network to reduce the task computing delay and minimize the control overhead in the network.
In \cite{w13}, Liu \textit{et al.} considered the dynamic requirements and resource constraints in the two-layer VFC architecture and proposed a task offloading strategy to maximize the task service rate.
In \cite{w14}, Wang \textit{et al.} proposed an online learning-based task offloading algorithm in the VFC to minimize the task offloading latency.
In \cite{w15}, Zhu \textit{et al.} proposed an event-triggered dynamic task allocation framework in the VFC by using linear programming-based optimization and binary particle swarm optimization to minimize average service latency.
In \cite{w16}, Zhao \textit{et al.} proposed a contract-based incentive mechanism that combines resource contribution and resource utilization in the VFC to improve the quality of service of vehicles. Moreover, they proposed a task offloading strategy based on the queuing model to enhance the performance of task offloading and resource allocation.
In \cite{w17}, Tang \textit{et al.} considered the task deadline and proposed an offloading strategy in the VFC to enhance the offloading efficiency.
In \cite{w18}, Yang \textit{et al.} considered the effective time of tasks in the VFC at the crossroads and proposed a low-cost offloading strategy to enhance the offloading efficiency among vehicles.
In \cite{w19}, Mourad \textit{et al.} considered the intrusion detection tasks in the VFC and proposed a task offloading strategy to minimize the computation execution time and energy consumption and meanwhile maximize the offloading survivability.
In \cite{w20}, Liu \textit{et al.} considered a two-layer VFC and proposed an adaptive task offloading mechanism to get the maximal  tasks' completion ratio.
In \cite{w21}, Cho \textit{et al.} adopted the adversarial multi-armed bandit theory to propose an adversarial online learning algorithm with bandit feedback, which aimed to optimize the selection of vehicles in the VFC to minimize the offloading service costs including the delay and energy.
In \cite{w22}, Wu \textit{et al.} considered directional vehicle mobility to propose a network model to minimize the average response time of the tasks. They further chose the neighboring vehicles to assist to process task based on a greedy algorithm.
In \cite{w24}, Huang \textit{et al.} jointly considered the task type as well as the velocity of vehicle to propose the task offloading and resource allocation strategy, which decreases the vehicles' energy cost and increases the income of the vehicles considering the delay constraint.
In \cite{w25}, Alam \textit{et al.} developed a three-layer generic decentralized cooperative VFC system to overcome the lack of robustness against the high mobility environment.
In \cite{w26}, Fan \textit{et al.} proposed a joint task offloading and resource allocation strategy for a VFC including devices and vehicles covered by a base station.
In \cite{w27}, Chen \textit{et al.} focused on a energy-powered multi-server VFC system with cybertwin, where vehicles send the current status of network and unprocessed vehicular application tasks to the macro base station to get the efficient resources allocation.
In \cite{w28}, Chen \textit{et al.} proposed a distributed multi-hop task offloading  model to get a low delay for the task execution in the VFC.
In \cite{w29}, Lin \textit{et al.} proposed an online task offloading strategy for the unknown dynamics heterogeneous VFC environment based on the online clustering of bandits (CAB) approaches to minimize the expectation of total offloading energy consumption while satisfy stringent delay requirements by learning the relationship between historical observations and rewards.
In \cite{w30}, Liu \textit{et al.} considered the vehicle mobility and proposed a task offloading strategy by exploiting the computation resources of vehicles in the VFC to get the minimized weighted sum of execution delay as well as computation cost.
In \cite{w31}, Ma \textit{et al.} proposed a traffic routing-based computation offloading strategy in the cybertwin-driven VFC for vehicle-to-everything applications to reduce latency. However, these works have not considered the platooning scenario.

\textcolor{black}{The works mentioned above show that the strategy of task offloading in VFC has been studied intensively, but for task offloading in the VFC-assisted platoon, to mitigate the limited computation resource, has not been studied. This motivates us to conduct this work. Different from the existing works about task offloading in VFC or platoons, we jointly take into account several factors, i.e., the heterogeneous computation resources of vehicles in the platoon, the random arrival and departure of task, the random arrival and departure of the vehicles in the VFC, and the 802.11p DCF mechanism. Moreover, we adopt the SMDP to model the task offloading process and design the SMDP framework including state, action, reward and transition probability.}

\section{System Model}\label{system}
In this section, we first describe \textcolor{black}{the VFC-assisted system}, and then briefly review the IEEE 802.11p DCF mechanism.

\subsubsection{\textcolor{black}{The VFC-assisted platoon}}

Consider $N$ autonomous vehicles in a platoon driving on a lane with the same velocity and a VFC consisting of the personal vehicles within the communication range of the head vehicle of the platoon. The $i$-th vehicle in the platoon is ${V_i}$ $(i = 1, \ldots ,N)$, where ${V_1}$ is the leader vehicle and ${V_2}, {V_3}, \ldots, {V_N}$ are member vehicles. \textcolor{black}{The computing rate of ${V_i}$, i.e., central processing unit (CPU) cycles per second,} is denoted as ${f_i}(i = 1, \ldots ,N)$. The resource of each vehicle in the VFC is virtualized as a resource unit (RU) \cite{s1}. Each RU's computing rate is the same and denoted as ${f_v}$. We consider the computing rate of an autonomous vehicle in the platoon is larger than the computing rate of a RU, thus ${f_i}>{f_v}$. The arrival and departure rate of vehicles in the VFC are denoted as ${\lambda _v}$ and ${\mu _v}$, respectively. The maximum number of vehicles in the VFC system is $K\left( {K \ge 1} \right)$.

\begin{figure*}[htbp]
\centering
\includegraphics[width=\linewidth, scale=1.00]{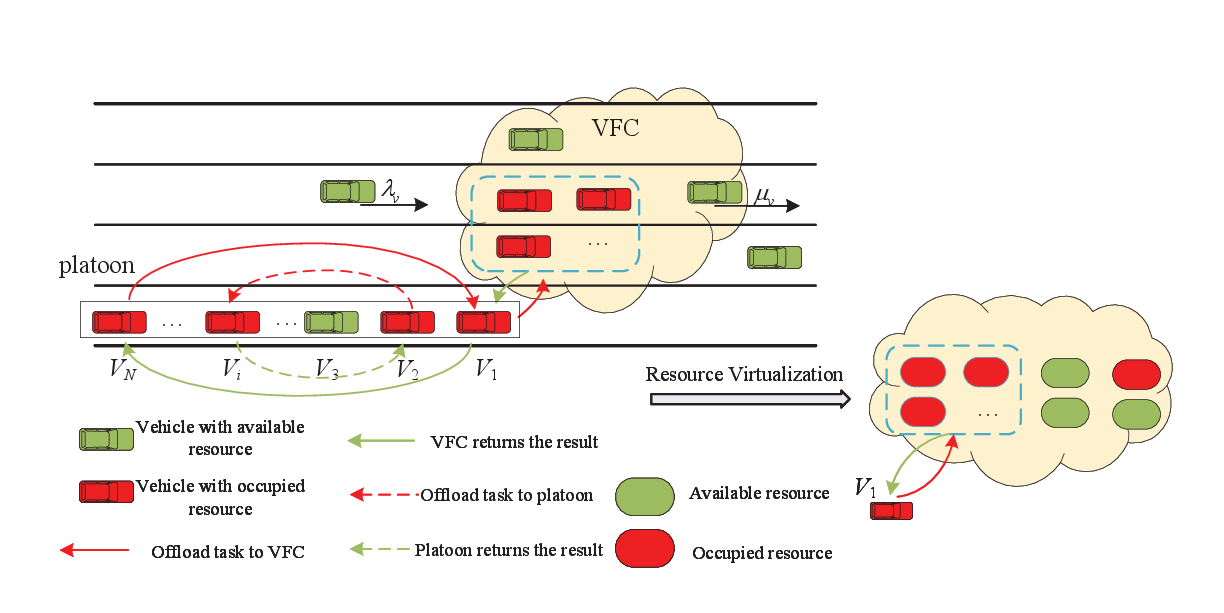}
\caption{\textcolor{black}{The task offloading in the VFC-assisted platoon system}}
\label{fig1}
\vspace{-0.5cm}
\end{figure*}

\begin{figure*}[htbp]
\centering
\includegraphics[width=\linewidth, scale=1.00]{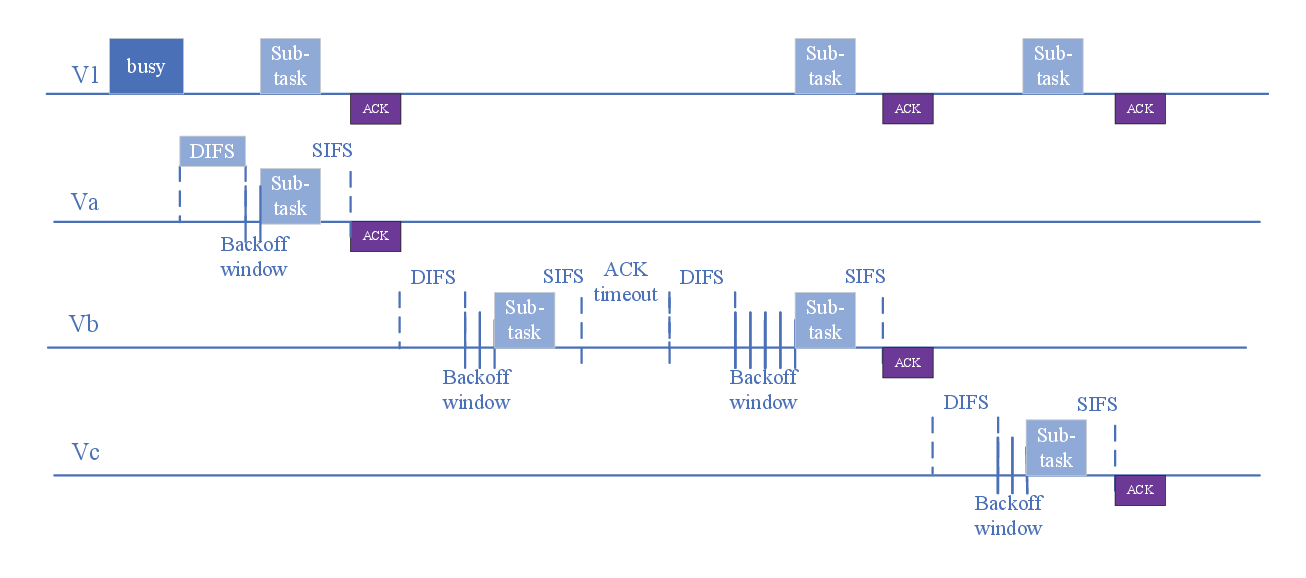}
\caption{Transmitting process of 802.11p DCF}
\label{fig2}
\vspace{-0.5cm}
\end{figure*}

The tasks arrive at each vehicle in the platoon according to the \textcolor{black}{Poisson distribution} with arrival rate ${\lambda _p}$. When a task arrives at a vehicle in the platoon, the system will make a decision to offload the task to the platoon, VFC, or to discard the task according to the available resources in the platoon and VFC. We consider the following two most general cases, (1) the task can be offloaded to only one vehicle in the platoon due to the sufficient computing rate of each vehicle in the platoon and (2) the task can be offloaded to multiple RUs in the VFC due to the limited computing rate of each RU. Specifically, if the available resources in the platoon are sufficient, the vehicle with the arrival task will request to offload the task to the platoon. Otherwise, the system will check if the VFC has sufficient computation resource and offload the task to it instead.
In this case, the requested vehicle first transmits the task to the leader vehicle of the platoon, \textcolor{black}{then the system determines the vehicles in the VFC with available resources to process. The leader vehicle divides the task into the corresponding number of subtasks, and transmits the subtasks to the corresponding vehicles in the VFC one by one. After that each vehicle processes the assigned subtasks to him. Once all the subtasks are executed, the result is sent back to the leader vehicle and further forwarded to the requested vehicle. Note that after the system determines the VFC vehicles, they will not be occupied by other subtasks until all of them have finished the processing.} Since the result is of small size, the backhaul delay is significantly small and thus can be neglected \cite{D1}. If the available resources are scarce in the platoon and VFC, the system will discard the task.
\textcolor{black}{Fig. \ref{fig1} shows an example for the offloading process including task offloading to platoon and to VFC. The task offloading to platoon in Fig. \ref{fig1} is described as follows. When a task arrives at $V2$ in the platoon, and another vehicle $V_i$ in the same platoon has available resource, $V2$ offloads the task to $V_i$ for processing. After $V_i$ completes processing the task, it returns the result back to $V2$. The task offloading to VFC in Fig. \ref{fig1} is described as follows. When another task arrives at $V_N$ and all the sources of the vehicles in the platoon are occupied. $V_N$ first offloads the task to the leader vehicle $V_1$ in the platoon. Then $V_1$ offloads the task to three RUs in the VFC for processing. After processing, the result is returned to $V_1$ and then returned back to $V_N$.}

In the offloading process, it is assumed that all vehicles in the platoon and VFC communicate with each other within one-hop communication area by adopting the 802.11p DCF mechanism. We will introduce the process of the 802.11p DCF mechanism.

\subsubsection{802.11p DCF mechanism}

\textcolor{black}{At the beginning, each vehicle sets the minimum contention window (${W_{\min }}$) based on 802.11p DCF mechanism \cite{802.11p}.} When a vehicle transmits a task to ${N_v}$ $\left( {{N_v} = 1,2,3...} \right)$ vehicles, the vehicle will partition the task into ${N_v}$ sub-tasks. When the vehicle transmits the one sub-task to an allocated vehicle, it first detects the state of the channel. If the channel keeps idle for distributed inter-frame space (DIFS), the vehicle will transmit the sub-task. Otherwise, if the channel is busy, the vehicle will continue to detect the channel state until the channel keeps idle for DIFS. Then a back-off process is initialized to transmit the sub-task. Specifically, the vehicle randomly selects a natural number within $[0,W - 1]$ for the backoff counter, where $W$ is the contention window and is initialized as ${W_{\min }}$ (i.e., the minimum contention window). The value of the backoff counter is decremented by one if the vehicle detects the idle state in a time slot. The vehicle will transmit the sub-task if the backoff counter is decreased to zero.

The sub-task is transmitted successfully if the vehicle receives an acknowledgment (ACK) after a short inter-frame space (SIFS). If the vehicle has not received an ACK after a specific ACK time out interval, the transmission fails. Then the vehicle will initialize another backoff process to retransmit the sub-task, where the contention window is doubled. If the number of retransmissions $k$ does not reach the retransmission limit $m$, the contention window $W$ will be doubled to ${2^k}{W_{\min }}$ and a new back-off process is initialized for retransmission. Otherwise, if the number of retransmissions $k$ reaches the maximum number $m$, the vehicle will discard the sub-task and reset the contention window as ${W_{\min }}$ to transmit the next sub-task. The following sub-tasks are transmitted sequentially according to the above procedure until all sub-tasks are transmitted successfully. Fig. 2 shows the process that the leader
vehicle V1 transmits three sub-tasks to three vehicles (denoted
as ${V_a}$, ${V_b}$ and ${V_c}$) in the VFC system.

\section{SMDP Framework}
In this section, we adopt the SMDP to model the task offloading process \textcolor{black}{in the VFC-assisted platoon system}, where the system learns the offloading decision by interacting with the environment in continuous time steps. The duration of each time step follows an exponential distribution. In each step, the system observes the state and then makes an action according to a policy. Afterwards, the system receives a reward which is related to the offloading delay to evaluate the benefit of the system under the state and action. The state transits to the next one according to the state transition probability. After that the system starts the next time step and repeats the above procedures. The target of the SMDP is to find the optimal policy to maximize the long-term reward.
Next, we construct the SMDP framework which includes the state set, action set, reward function and state transition probability. For clearity, the notations in this paper are listed in Table \ref{tab1}.
\begin{table*}
\caption{Notations used in this paper}
\label{tab1}
\renewcommand{\arraystretch}{1.1}
\centering
	\begin{tabular}{|p{2cm}<{\centering}|p{5.9cm}|p{2cm}<{\centering}|p{5.9cm}|}	\hline
		\textbf{Notation} &\textbf{Description} &\textbf{Notation} &\textbf{Description}\\
\hline
$N$ & Number of vehicles in the platoon. & $k$ & Number of retransmission. \\
\hline
${V_i}$ & The $i$-th vehicle in the platoon. & ${\lambda _v}/{\mu _v}$ & \multicolumn{1}{m{6cm}|}{Vehicles' arriving /departing rate.} \\
\hline
${\lambda _p}$ & Task \textcolor{black}{arrival} rate. & $K$ & Maximum number of RUs. \\
\hline
$d$ & Required CPU cycles to process each task. & $f_i$ & \multicolumn{1}{m{6cm}|}{Computing rate of vehicle ${V_i}$ (CPU cycles per second).} \\
\hline
${f_v}$ & Computing rate of each RU. & ${W_{\min}}$ & Minimum contention window. \\
\hline
$m$ & Maximum number of retransmissions. & $A$ & A task arrives at the system. \\
\hline
$D_i$ & \multicolumn{1}{m{6cm}|}{A task which occupies the resource of ${V_i}$ departs the system.} & ${L_j}$ & \multicolumn{1}{m{6cm}|}{A task which occupies $j$ RUs departs the system.}\\
\hline
${N_R}$ & \multicolumn{1}{m{6cm}|}{Maximum number of RUs that each task can occupy.} & ${F_ + }_1/{F_{ - 1}}$ & A vehicle arrives at/ departs the VFC. \\
\hline
$b$ & \multicolumn{1}{m{6cm}|}{The system discards the task.} & $b_i^p$ & Vehicle ${V_i}$ is allocated to process the task. \\
\hline
$b_j^{vf}$ & \multicolumn{1}{m{6cm}|}{The system allocates $j$ RUs in the VFC to process the task.} & ${n_i}$ & A binary to indicate whether the resource of ${V_i}$ is occupied.\\
\hline
${B_j}$ & Number of the tasks that occupy $j$ RUs. & $P$ & State transition probability of the system. \\
\hline
$M$ & Total number of RUs. & ${E_l}$ & Delay that processing the task locally.\\
\hline
${T_p}$ & \multicolumn{1}{m{6cm}|}{Transmitting delay that a vehicle transmits the task to another vehicle in the platoon.} & $T_j^{vf}$ & Transmitting delay that the leader vehicle transmits the task to $j$ RUs. \\
\hline
$\eta $ & \multicolumn{1}{m{6cm}|}{Saved price of each time unit.} & $\theta $ & \multicolumn{1}{m{6cm}|}{Number of the sub-tasks need to be transmitted.} \\
\hline
${E_{tr}}$ & \multicolumn{1}{m{6cm}|}{Average number of time slots to transmit a task.} & ${T_{slot}}$ & Average length of a time slot.\\
\hline
${N_{tr}}$ & \multicolumn{1}{m{6cm}|}{Number of vehicles in the one-hop communication area when a vehicle transmits a task.} & ${q_{idle}}/{q_s}/{q_c}$ & Probability of idle/successful transmission/collision.\\
\hline
${T_{idle}}/{T_s}/{T_c}$ & \multicolumn{1}{m{6cm}|}{Duration of idle/successful transmission/collision.} & $E\left[ P \right]$ & Length of a task.\\
\hline
$\delta$ & Propagation delay. & $\omega$ & \multicolumn{1}{m{6cm}|}{Transmission probability of a vehicle in a time slot.}\\
\hline
$q$ &  \multicolumn{1}{m{6cm}|}{Collision probability when a vehicle is transmitting.} & $\zeta$ & \multicolumn{1}{m{6cm}|}{Punishment of the system discarding the task/the vehicle which is processing the task departs the system.} \\
\hline
$\alpha$ & Continuous-time discount factor. & & \\
\hline
\end{tabular}
\vspace{-0.7cm}
\end{table*}
\subsection{State set}
The states reflect the resource conditions of the system, which is influenced by the events including a task's arrival and departure, a vehicle's arrival and departure in the VFC. Hence, we formulate the states as the resource conditions when different events occur. The state set $\mathbb{S}$ is formulated as
\begin{equation}
\begin{aligned}
\mathbb {S} = \{ {n_1},\ldots ,{n_N},{B_1},\ldots ,{B_{{N_R}}},M,e\},
\label{eq1}
\end{aligned}
\end{equation}
where ${n_i}$  ($i = 1, \ldots ,N$) is a binary to indicate whether the resource of ${V_i}$ is occupied or not,  i.e., ${n_i} = 1$ indicates that the resource of ${V_i}$ is occupied and ${n_i} = 0$ indicates that the resource of ${V_i}$ is available. ${B_j}$ ($j = 1, \ldots ,{N_R}$) is the number of the tasks that occupy $j$ RUs, ${N_R}$ is the maximum number of RUs that each task can occupy, $M$ is the total number of RUs in the VFC. \textcolor{black}{Thus, the total number of occupied RUs is calculated as $\sum\nolimits_{r = 1}^{{N_R}} {r{B_r}}$, where $r$ is the number of RUs a task occupies, and ${B_r}$ is the number of the tasks that occupy $r$ RUs. Then ${r{B_r}}$ represents the number of the occupied RUs for the tasks occupied $r$ RUs. The total number of occupied RUs should not exceed the total number of RUs in VFC, i.e., $\sum\nolimits_{r = 1}^{{N_R}} {r{B_r}}  \le M$.} $e \in \{ A,{D_1}, \ldots ,{D_N},{L_1}, \ldots ,{L_{{N_R}}},{F_{ + 1}},{F_{ - 1}}\} $ is an event,  where $A$ denotes that a task arrives at the system, ${D_i}$ denotes that a task occupying the resource of ${V_i}$ leaves the system, \textcolor{black}{${L_j}$ denotes that a task occupying $j$ RUs in the VFC leaves the system,} ${F_{ + 1}}$ denotes that a vehicle arrives at the VFC, ${F_{ - 1}}$ denotes that a vehicle leaves the VFC.

\subsection{Action set}
The actions reflect the offloading decisions of the system, depending on different events. Specifically, when there is a task arriving, the system will make an offloading decision such as offloading to the platoon, offloading to the VFC or discarding the task, while the system will make no decision when other events occur. Thus, the action set $\mathbb{A}$ is formulated as
\begin{equation}
\begin{aligned}
\begin{array}{l}
\mathbb{A} =
\left\{ \begin{array}{l}
\{ b,b_1^p, \ldots ,b_N^p,b_1^{vf}, \ldots ,b_{{N_R}}^{vf}\} , e = A \\
\\
\{{b_{ - 1}}\},e\in\{{D_1},\ldots,{D_N},{L_1},\ldots,{L_{{N_R}}},{F_{ + 1}},{F_{ - 1}}\}
\end{array}\!\!, \right.
\end{array}
\end{aligned}
\label{eq2}
\end{equation}
\textcolor{black}{where $b_i^p$ indicates that vehicle ${V_i}$ in the platoon is allocated to process the offloaded task, $b_j^{vf}$ indicates that the system offloads the task to the VFC and allocates total $j$ RUs in the VFC to process the task, $b$ indicates that the system discards the task. ${b_{ - 1}}$ indicates that the system takes no action in the current moment.}

\subsection{Reward function}
The reward function reflects the benefit which is related to the offloading delay under different states and actions. The system will transit from the current state $s$ to the next state when an action $a$ is taken, thus it gets an income $U(s,a)$ when taking action $a$ under state $s$, and incurs a cost $G(s,a)$ during the duration from state $s$ to the next state. Hence, the reward function under state $s$ and action $a$ is calculated as
\begin{equation}
R(s,a) = U(s,a) - G(s,a).
\label{eq3}
\end{equation}
Next, we will formulate $U(s,a)$ and $G(s,a)$, respectively.
\subsubsection{Income}
The state changes when an event occurs. Hence, we will formulate the  income $U(s,a)$ under the following situations.
\begin{itemize}
\item[(a)] $a = b_i^p, e=A$: When there is a task arriving and the available resources of the platoon are sufficient, the system makes an offloading decision to offload the task to vehicle ${V_i}$ in the platoon. In this case, it has a lower offloading delay than being processed locally. The income is the saved delay. Hence, the income is formulated as  $\eta \left( {{E_l} - {T_p} - {d \mathord{\left/{\vphantom {d {{f_i}}}} \right.\kern-\nulldelimiterspace} {{f_i}}}} \right)$, \textcolor{black}{where $\eta $ is a constant to denote the saved price of each time unit \cite{s1},} $E_l$ is the delay for processing the task locally,
the transmitting delay ${T_p}$ means the duration that a vehicle transmits the task to another vehicle in the platoon, $d$ is the required CPU cycles to process each task, thus the processing delay of vehicle $V_i$ is ${{d} \mathord{\left/{\vphantom {{d} {f_i}}} \right. \kern-\nulldelimiterspace} f_i}$.
\item[(b)]$a = b_j^{vf}, e=A$: When there is a task arriving, VFC may have sufficient resources but not for the platoon, the system allocates $j$ RUs to process the task. Thus the task will be first transmitted to the leader vehicle, which then transmits the task to $j$ RUs for processing. Hence, task offloading delay is formulated as $\eta ({E_l} - {T_p} - T_j^{vf} - {d \mathord{\left/ {\vphantom {d {j{f_v}}}} \right. \kern-\nulldelimiterspace} {(j{f_v})}})$, where $T_j^{vf}$ is the transmitting delay that the leader vehicle transmits the task to $j$ RUs, and ${{d} \mathord{\left/{\vphantom {{d} j{f_v}}} \right.\kern-\nulldelimiterspace} (j{f_v})}$ is the processing delay of $j$ RUs.
\item[(c)]$a = b, e = A$: When there is a task arriving and the available resources of both platoon and VFC are insufficient, the system will make a decision to discard the task. This decision is detrimental because the vehicle cannot receive the result, thus the system receives a punishment $- \zeta $.
\item[(d)]$a = {b_{ - 1}, e \in \{ {D_1}, \ldots ,{D_N},{L_1}, \ldots ,{L_{{N_R}}},{F_{ + 1}}\}}$: When a task departs the system or there is a vehicle arriving at the VFC, the system do not take action. In this situation, the income of the system is $0$.
\item[(e)]$a = {b_{ - 1}}, e = {F_{ - 1}}, \sum\nolimits_{r = 1}^{{N_R}}{r{B_r}}<M$: When there is a vehicle departing the VFC and the VFC has the available RUs, the system takes no action and the income is also $0$.
\item[(f)]$a = {b_{ - 1}}, e = {F_{ - 1}}, \sum\nolimits_{r = 1}^{{N_R}}{r{B_r}}=M$: When a vehicle departs the VFC and there is no available RU in the VFC, the system also does not take action, but the departing vehicle will interrupt its task processing. Hence, the system receives a punishment $ - \zeta $.
\end{itemize}

Conclusively, the income of system is expressed as
\begin{equation}
\begin{aligned}
\begin{array}{l}
U(s,a) = \\
\left\{ \begin{array}{l}
\eta ({E_l} - {T_p} - \frac{d}{{{f_i}}}), \qquad\qquad\quad a = b_i^p,e = A\\
\eta ({E_l} - {T_p}-T_j^{vf}-\frac{d}{{j{f_v}}}), \!\!\qquad a = b_j^{vf}, e = A\\
 - \zeta, \qquad a = b, e = A\\
0, \qquad\quad \!a={b_{ - 1}},e \in \{ {D_1},\ldots ,{D_N},{L_1},\ldots ,{L_{{N_R}}},{F_{+1}}\}\\
0, \qquad\quad \!a={b_{ - 1}}, e={F_{ - 1}}, \sum\nolimits_{r = 1}^{{N_R}} {r{B_r}}< M\\
 - \zeta, \qquad a = {b_{ - 1}}, e = {F_{ - 1}}, \sum\nolimits_{r = 1}^{{N_R}}{r{B_r}}= M
\end{array}\!\!\!. \right.
\end{array}
\label{eq4}
\end{aligned}
\end{equation}

In  \eqref{eq4}, as ${T_p}$ and $T_j^{vf}$ depend on the 802.11p DCF mechanism, we will derive ${T_p}$ and $T_j^{vf}$ accordingly. Let $\theta $ be the number of transmitted sub-tasks, ${T_{slot}}$ be the average length of a time slot, ${E_{tr}}$ be the average number of time slots to transmit a task, thus the transmitting delay ${T_{tr}}$ is calculated as
\begin{equation}
{T_{tr}} = \theta {T_{slot}}{E_{tr}}.
\label{eq5}
\end{equation}
When a vehicle transmits the task to another vehicle in the platoon, the task is not divided and $\theta = 1$. In this case, ${T_{p}}={T_{slot}}{E_{tr}}$. When the leader vehicle of the platoon transmits a task to $j$ RUs, the task is partitioned into $j$ sub-tasks and $\theta = j$, and thus we can get $T_j^{vf}=j{T_{slot}}{E_{tr}}$.


Next, we will further derive ${T_{slot}}$ and ${E_{tr}}$, respectively. For the 802.11p DCF mechanism, a time slot may be at different statuses, such as successful transmission, collision or idle. Thus the average duration of a time slot can be expressed as
\begin{equation}
{T_{slot}} = {q_{idle}}{T_{idle}} + {q_s}{T_s} + {q_c}{T_c},
\label{eq6}
\end{equation}
where ${q_{idle}}$, ${q_s}$ and ${q_c}$ are the probabilities of idle, successful transmission and collision, respectively. ${T_{idle}}$, ${T_s}$ and ${T_c}$ indicates the durations of idle, successful transmission and collision, respectively.


Based on the 802.11p DCF mechanism\cite{D2,wu14,wu15}, ${T_s}$ and ${T_c}$ are calculated, respectively, as
\begin{equation}
\left\{ \begin{array}{l}
{T_s} = H_{h}+{{E[P]} \mathord{\left/
 {\vphantom {{E[P]} \theta }} \right.
 \kern-\nulldelimiterspace} \theta } + SIFS + \delta  + ACK + DIFS + \delta  \\
{T_c}= H_{h}{\rm{ + }}{{E[P]} \mathord{\left/
 {\vphantom {{E[P]} \theta }} \right.
 \kern-\nulldelimiterspace} \theta } + SIFS + \delta + ACK_{timeout}
\end{array} \right.\!\!,
\label{eq7}
\end{equation}
where $H_{h}$ denotes the packet header's length, $E[P]$ is a task's length, $\delta$ is the propagation delay, $SIFS$, $ACK$ and $DIFS$ are the length of SIFS, ACK and DIFS, respectively, and $ACK_{timeout}$ is the ACK time out interval's length.

Next, ${q_{idle}}$, ${q_s}$ and ${q_c}$ are further derived. A time slot is idle when no other vehicles are transmitting in the one-hop communication range, which facilitates one vehicle to execute a successful transmission. Thus we have
\begin{equation}
{q_{idle}} = {(1 - \omega )^{{N_{tr}}}},
\label{eq8}
\end{equation}
\begin{equation}
{q_s} = \omega {N_{tr}}{(1 - \omega )^{{N_{tr}} - 1}},\
\label{eq9}
\end{equation}
\begin{equation}
{q_c} = 1 - {q_{idle}} - {q_s},
\label{eq10}
\end{equation}
where $\omega$ is the transmission probability of a vehicle in a time slot and ${N_{tr}}$ is the number of vehicles in the one-hop communication area when a vehicle transmits a task. Note that vehicles in the platoon communicate with each other by one-hop communication over a common channel and the head vehicle communicates with the vehicles in the VFC through one-hop communication over another channel. Thus ${N_{tr}}$ equals to $N$ when a vehicle is transmitting the task to another vehicle in the platoon while ${N_{tr}}$ becomes $M + 1$ when the leader vehicle is transmitting the task to vehicles in the VFC.

In Eqs. \eqref{eq8}-\eqref{eq10}, $\omega$ can be calculated by solving the following equations \cite{D3}, i.e.,

\begin{equation}
q = 1 - {(1 - \omega )^{{N_{tr}} - 1}},
\label{eq11}
\end{equation}
\begin{equation}
\omega  = \frac{{2(1 - 2q)}}{{(1 - 2q)({W_{\min }} + 1) + q{W_{\min }}(1 - {{(2q)}^m})}},
\label{eq12}
\end{equation}
where $q$ is the collision probability when a vehicle is transmitting and $W_{\min}$ is the minimum contention window.

Moreover, according to \cite{D4}, ${E_{tr}}$ in  \eqref{eq5} can be calculated as

\begin{equation}
\begin{split}
{E}_{tr}=&\frac{1-(m+2){{q}^{m+1}}+(m+1){{q}^{m+2}}}{2(1-q)}+\\
&\frac{(1-q)[1-{{(2q)}^{m+1}}]{{W}_{\min }}}{1-2q}-\frac{(1-{{q}^{m+1}}){{W}_{\min }}}{2}+\\
&\frac{{{q}^{m+1}}}{2}[ m+1+({{2}^{m+1}}-1){{W}_{\min }}+\\
&\frac{(2-q)({{2}^{m}}{{W}_{\min }}+1)}{1-q}]
\end{split}
\label{eq13}.
\end{equation}

As derived above, given $W_{\min}$, one can calculate ${T_{slot}}$ and ${E_{tr}}$ according to Eqs. \eqref{eq6}-\eqref{eq13}, and ${T_p}$ and $T_j^{vf}$ according to  \eqref{eq5}, then we can get the income
according to  \eqref{eq4}.


\subsubsection{Cost}
 We consider the long-term expected discounted cost during the time step as the cost $G(s,a)$. Since the duration of each time step is assumed to follow an exponential distribution with parameter $\alpha$, $G(s,a)$ is calculated as \cite{D7}:
\begin{equation}
\begin{aligned}
G\left( {s,a} \right) &= C\left( {s,a} \right)E_s^a\left\{ {\int\limits_0^\tau  {{e^{ - \alpha t}}dt} } \right\}\\
 &= C\left( {s,a} \right)E_s^a\left\{ {\frac{{1 - {e^{ - \alpha \tau }}}}{\alpha }} \right\}\\
 &= \frac{{C(s,a)}}{{\alpha  + \beta (s,a)}}
\label{eq16}
\end{aligned},
\end{equation}
where $\alpha$ is the continuous-time discount factor, $C\left({s,a}\right)$ is the expected service time's cost rate under state $s$ and action $a$, which is calculated as the number of vehicles that are occupied in the system, i.e.,
\begin{equation}
C(s,a) = \sum\nolimits_{k = 1}^{N} {{n_k}}  + \sum\nolimits_{r = 1}^{{N_R}} {r{B_r}}.
\label{eq14}
\end{equation}
Meanwhile, $\beta (s,a)$ is the sum of the rate of all events under state $s$ and action $a$. In the system, the rate that \textcolor{black}{the vehicles arrive} at the system is $\lambda _v$, the rate that \textcolor{black}{the vehicles depart} the system is $\mu _v$, and the rate that tasks arrive at the system is $N{\lambda _p}$. In addition, the number of the tasks occupying vehicle ${V_i}$ and $j$ RUs are $n_i$ and $B_j$, respectively, thus the rates that the departure tasks depart the system after being processed at vehicles in the platoon and VFC can be expressed as $\sum\nolimits_{r = 1}^{N} {{n_k}\frac{{{f_k}}}{d}}$ and $\sum\nolimits_{r = 1}^{{N_R}} r{{B_r}\frac{{{f_v}}}{d}}$, respectively. Since the state is changed when an event occurs, we will further analyze $\beta (s,a)$ under different actions for different events.

\begin{itemize}
\item[(a)] $a = b_i^p, e=A$: When there is a task arriving, the system offloads the task to $V_i$. Thus the rate that the tasks occupying the vehicles in the platoon depart the system becomes $\sum\nolimits_{r = 1}^{N} {{n_k}\frac{{{f_k}}}{d}}+ \frac{{{f_i}}}{d}$.

\item[(b)] $a = b_j^{vf}, e=A$: When there is a task arriving, system offloads the task to $j$ vehicles in VFC. Thus the rate that the tasks depart the system after being processed at vehicles of the VFC becomes $\sum\nolimits_{r = 1}^{{N_R}} r{{B_r}\frac{{{f_v}}}{d}}+j\frac{{{f_v}}}{d}$.

\item[(c)] $a = b_{-1}, e=D_i$: When there is a task occupying vehicle $V_i$ in the platoon, i.e., $n_i=1$, departs the system, the system makes no action. Thus the rate that the tasks occupying the vehicles in the platoon depart the system becomes $\sum\nolimits_{r = 1}^{N} {{n_k}\frac{{{f_k}}}{d}}- \frac{{{f_i}}}{d}$.

\item[(d)] $a = b_{-1}, e=L_j$: When a task occupying $r$ vehicles in the VFC departs from the system, the system makes no action. Thus the corresponding event rate becomes $\sum\nolimits_{r = 1}^{{N_R}} r{{B_r}\frac{{{f_v}}}{d}}-r\frac{{{f_v}}}{d}$.

\item[(e)] $a = b_{-1}, e=F_{+1}/F_{-1}$: When there is a vehicle arriving/departing the system, the system makes no action. Thus the rate of each event does not change.
\end{itemize}

Based on analysis above, $\beta (s,a)$ is summarized as:
\begin{equation}
\begin{array}{l}
\beta (s,a) = \\
\left\{ \begin{array}{l}
N{\lambda _p} +{\lambda _v} + {\mu _v} + \sum\nolimits_{r = 1}^{N} {{n_k}\frac{{{f_k}}}{d}} + \sum\nolimits_{r = 1}^{{N_R}} {{B_r}r\frac{{{f_v}}}{d}}  + \frac{{{f_i}}}{d},\\
e = A,a = b_i^p\\
N{\lambda _p} + {\lambda _v} + {\mu _v} + \sum\nolimits_{k = 1}^N {{n_k}\frac{{{f_k}}}{d}}  + \sum\nolimits_{r = 1}^{{N_R}} {{B_r}r\frac{{{f_v}}}{d}}  + j\frac{{{f_v}}}{d},\\
e = A,a = b_j^{vf}\\
N{\lambda _p} + {\lambda _v} + {\mu _v} + \sum\nolimits_{k = 1}^N {{n_k}\frac{{{f_k}}}{d}}  + \sum\nolimits_{r = 1}^{{N_R}} {{B_r}r\frac{{{f_v}}}{d}}  - \frac{{{f_i}}}{d},\\
e = {D_i},a = {b_{ - 1}}\\
N{\lambda _p} + {\lambda _v} + {\mu _v} + \sum\nolimits_{k = 1}^N {{n_k}\frac{{{f_k}}}{d}}  + \sum\nolimits_{r = 1}^{{N_R}} {{B_r}r\frac{{{f_v}}}{d}}  - j\frac{{{f_v}}}{d},\\
e = {L_j},a = {b_{ - 1}}\\
N{\lambda _p} + {\lambda _v} + {\mu _v} + \sum\nolimits_{k = 1}^N {{n_k}\frac{{{f_k}}}{d}}  + \sum\nolimits_{r = 1}^{{N_R}} {{B_r}r\frac{{{f_v}}}{d}} .\\
e = {F_{ + 1}}/{F_{ - 1}},a = {b_{ - 1}}
\end{array} \right.
\end{array}
\label{eq18}
\end{equation}

Finally, $G(s,a)$ can be calculated by substituting Eqs. \eqref{eq14} and \eqref{eq18} into  \eqref{eq16}.

\vspace{-0.2cm}
\subsection{State Transition Probability}

In this section, we will derive the state transition probability from the current state $s$ to the next state $s'$ after taking action $a$. Let $e'$ be the event at the next state $s'$, $P(s'|s,a)$ be the state transition probability from the current state $s$ to the next state $s'$ after taking action $a$.
The definition of $P(s'|s,a)$ is discussed in the following six cases.

When a task arrives under the current state and the task is offloaded to vehicle $V_i$ in the platoon, the number of the tasks occupying vehicle $V_i$ is incremented by one. In this case, six kinds of different rates are listed as follows: the rate that tasks arrive at the system is $N{\lambda _p}$, the rate that tasks occupying vehicle $V_i$ depart the system is $\left( {{n}_{i}}+1 \right)\frac{{{f}_{i}}}{d}$, the rate that the tasks occupying vehicle $u$ ($u \ne i$) depart from the system is ${{n_u}\frac{{{f_u}}}{d}}$, the rate that the tasks occupying $j$ RUs depart the system is $j{{B_j}\frac{{{f_v}}}{d}}$, the rate that vehicles arrive and depart the VFC are $\lambda_v$ and $\mu_v$, respectively.

Therefore, the corresponding $P(s'|s,a)$ are calculated respectively as follows:
\begin{enumerate}
\item[(a)] If the next event is a task arriving at the system, $P(s'|s,a)$ is calculated as $\frac{{N{\lambda _p}}}{{\beta (s,a)}}$.
\item[(b)] If the next event is that a task occupying vehicle $V_i$ leaves the system,  $P(s'|s,a)$ is calculated as $\frac{{({n_i} + 1)\frac{{{f_i}}}{d}}}{{\beta (s,a)}}$.
\item[(c)] If the next event is that a task occupying vehicle $u$ ($u \ne i$) leaves the system, $P(s'|s,a)$ is calculated as $\frac{{{n_u}\frac{{{f_u}}}{d}}}{{\beta (s,a)}}$.
\item[(d)] If the next event is that a task occupying $j$ vehicles in the VFC leaves the system, $P(s'|s,a)$ is calculated as $\frac{{{B_j}\frac{{j{f_v}}}{d}}}{{\beta (s,a)}}$.
\item[(e)] If the next event is a vehicle arriving at the VFC, $P(s'|s,a)$ is calculated as  $\frac{{{\lambda_v}}}{{\beta (s,a)}}$.
\item[(f)] If the next event is a vehicle leaving the VFC, $P(s'|s,a)$ is calculated as  $\frac{{{\mu_v}}}{{\beta (s,a)}}$.
\end{enumerate}

Based on the above analysis, when $s = ({n_1}, \ldots ,{n_N},{B_1}, \ldots ,{B_{{N_R}}},M,A)$ and $a = b_i^p$, the transition probability $P(s'|s,a)$ is given by

\begin{equation}
\begin{array}{l}
P(s'|s,a) = \\
\left\{ \begin{array}{l}
\frac{{N{\lambda _p}}}{{\beta (s,a)}},\\
s' = ({n_1}, \ldots ,{n_i} + 1, \ldots ,{n_N},{B_1}, \ldots ,{B_{{N_R}}},M,A)\\

\frac{{({n_i} + 1)\frac{{{f_i}}}{d}}}{{\beta (s,a)}},\\
s' = ({n_1}, \ldots ,{n_i} + 1, \ldots ,{n_N},{B_1}, \ldots ,{B_{{N_R}}},M,{D_i})\\

\frac{{{n_u}\frac{{{f_u}}}{d}}}{{\beta (s,a)}},u \ne i,\\
s' = ({n_1},\ldots ,{n_i} + 1, \ldots,{n_N},{B_1}, \ldots,{B_{{N_R}}},M,{D_u})\\

\frac{{{B_j}\frac{{j{f_v}}}{d}}}{{\beta (s,a)}},\\
s' = ({n_1}, \ldots ,{n_i} + 1, \ldots ,{n_N},{B_1}, \ldots ,{B_{{N_R}}},M,{L_j})\\

\frac{{{\lambda _v}}}{{\beta (s,a)}},\\
s' = ({n_1}, \ldots,{n_i}+1, \ldots,{n_N},{B_1}, \ldots,{B_{{N_R}}},M,{F_{ + 1}})\\

\frac{{{\mu _v}}}{{\beta (s,a)}},\\
s' =({n_1}, \ldots,{n_i} + 1, \ldots,{n_N},{B_1}, \ldots,{B_{{N_R}}},M,{F_{ - 1}})
\end{array} \right..
\end{array}
\label{eq19}
\end{equation}

Similarly, the transition probability $P(s'|s,a)$ under other events and actions is calculated as Eqs. \eqref{eq20}-\eqref{eq24}.

When $s = ({n_1}, \ldots ,{n_N},{B_1}, \ldots ,{B_{{N_R}}},M,A), a = b_j^{vf}$:
\begin{equation}
\begin{array}{l}
P(s'|s,a) = \\
\left\{ \begin{array}{l}
\frac{{N{\lambda _p}}}{{\beta (s,a)}},\\
s' = ({n_1}, \ldots ,{n_N},{B_1}, \ldots ,{B_j} + 1, \ldots ,{B_{{N_R}}},M,A)\\

\frac{{{n_i}\frac{{{f_i}}}{d}}}{{\beta (s,a)}},\\
s' = ({n_1}, \ldots,{n_N},{B_1}, \ldots,{B_j} + 1, \ldots ,{B_{{N_R}}},M,{D_i})\\

\frac{{({B_j}\! +\! 1)\frac{{j{f_v}}}{d}}}{{\beta (s,a)}},\\
s' = ({n_1}, \ldots ,{n_N},{B_1}, \ldots ,{B_j} + 1, \ldots,{B_{{N_R}}},M,{L_j})\\

\frac{{{B_c}\frac{{c{f_v}}}{d}}}{{\beta (s,a)}},c \ne j,\\
s' = ({n_1}, \ldots,{n_N},{B_1}, \ldots ,{B_j} + 1, \ldots,{B_{{N_R}}},M,{L_c})\\

\frac{{{\lambda _v}}}{{\beta (s,a)}},\\
s' = ({n_1}, \ldots ,{n_N},{B_1}, \ldots ,{B_j} + 1, \ldots,{B_{{N_R}}},M,{F_{ + 1}})\\

\frac{{{\mu _v}}}{{\beta (s,a)}},\\
s'  = ({n_1}, \ldots,{n_N},{B_1}, \ldots ,{B_j} + 1, \ldots,{B_{{N_R}}},M,{F_{ - 1}})
\end{array} \right..
\end{array}
\label{eq20}
\end{equation}
When $s = ({n_1}, \ldots ,{s_N},{B_1}, \ldots ,{B_{{N_R}}},M,{D_i}),a = {b_{ - 1}}$:
\begin{equation}
\begin{array}{l}
P(s'|s,a) = \\
\left\{ \begin{array}{l}
\frac{{N{\lambda _p}}}{{\beta (s,a)}},\\
s' = ({n_1}, \ldots ,{n_i} - 1, \ldots ,{n_N},{B_1}, \ldots ,{B_{{N_R}}},M,A)\\

\frac{{({n_i} - 1)\frac{{{f_i}}}{d}}}{{\beta (s,a)}},\\
s' = ({n_1}, \ldots ,{n_i} - 1, \ldots ,{n_N},{B_1}, \ldots ,{B_{{N_R}}},M,{D_i})\\

\frac{{{n_k}\frac{{{f_k}}}{d}}}{{\beta (s,a)}},k \ne i,\\
s' = ({n_1}, \ldots ,{n_i} - 1, \ldots ,{n_N},{B_1}, \ldots ,{B_{{N_R}}},M,{D_k})\\

\frac{{{B_j}j\frac{{{f_v}}}{d}}}{{\beta (s,a)}},\\
s' = ({n_1}, \ldots ,{n_i} - 1, \ldots ,{n_N},{B_1}, \ldots ,{B_{{N_R}}},M,{L_j})\\

\frac{{{\lambda _v}}}{{\beta (s,a)}},\\
s' = ({n_1}, \ldots ,{n_i} - 1, \ldots,{n_N},{B_1}, \ldots,{B_{{N_R}}},M,{F_{ + 1}})\\

\frac{{{\mu _v}}}{{\beta (s,a)}},\\
s' = ({n_1}, \ldots,{n_i} - 1, \ldots ,{n_N},{B_1},\ldots ,{B_{{N_R}}},M,{F_{ - 1}})
\end{array} \right..
\end{array}
\label{eq21}
\end{equation}
\vspace{-0.2cm}
When $s = ({n_1}, \ldots ,{n_N},{B_1}, \ldots ,{B_{{N_R}}},M,{L_j}),a = {b_{ - 1}}$:
\begin{equation}
\begin{array}{l}
P(s'|s,a) = \\
\left\{ \begin{array}{l}
\frac{{N{\lambda _p}}}{{\beta (s,a)}},\\
s' = ({n_1}, \ldots ,{n_N},{B_1}, \ldots ,{B_j} - 1, \ldots ,{B_{{N_R}}},M,A)\\

\frac{{{n_i}\frac{{{f_i}}}{d}}}{{\beta (s,a)}},\\
s' = ({n_1}, \ldots ,{n_N},{B_1}, \ldots ,{B_j} - 1, \ldots ,{B_{{N_R}}},M,{D_i})\\

\frac{{({B_j} - 1)j\frac{{{f_v}}}{d}}}{{\beta (s,a)}},\\
s' = ({n_1}, \ldots ,{n_N},{B_1}, \ldots ,{B_j} - 1, \ldots ,{B_{{N_R}}},M,{L_j})\\

\frac{{{B_c}c\frac{{{f_v}}}{d}}}{{\beta (s,a)}},c \ne j,\\
s' = ({n_1}, \ldots ,{n_N},{B_1}, \ldots ,{B_j} - 1, \ldots ,{B_{{N_R}}},M,{L_c})\\

\frac{{{\lambda _v}}}{{\beta (s,a)}},\\
s' = ({n_1}, \ldots ,{n_N},{B_1}, \ldots ,{B_j} - 1, \ldots ,{B_{{N_R}}},M,{F_{ + 1}})\\

\frac{{{\mu _v}}}{{\beta (s,a)}},\\
s' = ({n_1}, \ldots ,{n_N},{B_1},\ldots ,{B_j} - 1, \ldots ,{B_{{N_R}}},M,{F_{ - 1}})
\end{array} \right..
\end{array}
\label{eq22}
\end{equation}
When $s = ({n_1}, \ldots ,{n_N},{B_1}, \ldots ,{B_{{N_R}}},M,{F_{ + 1}}),a = {b_{ - 1}}$:
\begin{equation}
\begin{array}{l}
P(s'|s,a) = \\
\left\{ \begin{array}{l}
\frac{{N{\lambda _p}}}{{\beta (s,a)}},s' =({n_1},\ldots ,{n_N},{B_1}, \ldots ,{B_{{N_R}}},M + 1,A)\\

\frac{{{n_i}\frac{{{f_i}}}{d}}}{{\beta (s,a)}},s' = ({n_1}, \ldots ,{n_N},{B_1}, \ldots ,{B_{{N_R}}},M + 1,{D_i})\\

\frac{{{B_j}j\frac{{{f_v}}}{d}}}{{\beta (s,a)}},s' = ({n_1}, \ldots ,{n_N},{B_1}, \ldots ,{B_{{N_R}}},M + 1,{L_j})\\

\frac{{{\lambda _v}}}{{\beta (s,a)}},s' = ({n_1}, \ldots ,{n_N},{B_1}, \ldots ,{B_{{N_R}}},M + 1,{F_{ + 1}})\\

\frac{{{\mu _v}}}{{\beta (s,a)}},s' = ({n_1}, \ldots ,{n_N},{B_1}, \ldots ,{B_{{N_R}}},M + 1,{F_{ - 1}})
\end{array} \right..
\end{array}
\label{eq23}
\end{equation}
When $s = ({n_1}, \ldots ,{n_N},{B_1}, \ldots ,{B_{{N_R}}},M,{F_{ - 1}}),a = {b_{ - 1}}$:
\begin{equation}
\begin{array}{l}
P(s'|s,a) = \\
\left\{ \begin{array}{l}
\frac{{N{\lambda _p}}}{{\beta (s,a)}}, s' = ({n_1},\ldots ,{n_N},{B_1}, \ldots ,{B_{{N_R}}},M - 1,A)\\

\frac{{{n_i}\frac{{{f_i}}}{d}}}{{\beta (s,a)}},s' = ({n_1}, \ldots ,{n_N},{B_1}, \ldots ,{B_{{N_R}}},M - 1,{D_i})\\

\frac{{{B_j}j\frac{{{f_v}}}{d}}}{{\beta (s,a)}},s' = ({n_1}, \ldots ,{n_N},{B_1}, \ldots ,{B_{{N_R}}},M - 1,{L_j})\\

\frac{{{\lambda _v}}}{{\beta (s,a)}},s' = ({n_1}, \ldots ,{n_N},{B_1}, \ldots ,{B_{{N_R}}},M - 1,{F_{ + 1}})\\

\frac{{{\mu _v}}}{{\beta (s,a)}},s' = ({n_1}, \ldots ,{n_N},{B_1}, \ldots ,{B_{{N_R}}},M - 1,{F_{ - 1}})
\end{array} \right..
\end{array}
\label{eq24}
\end{equation}

\section{Solution}
In this section, we will employ an value iterative algorithm to derive the optimal policy, thus maximizing the long-term reward. Next, we will describe our algorithm in detail.

At the beginning, the iteration number and the value function of each state are initialized to zero. Then, for each iteration $l$, the maximum value function of each state under different actions is calculated based on the Bellman equation until the maximum value function of each state converges. Specifically, the continuous-time SMDP model is first transformed into the discrete-time SMDP model to enable the value iterative algorithm to solve it, where a normalization factor $y$ is introduced to normalize the reward, discount factor and state transition probability of the constructed continuous-time SMDP model \cite{D5}, i.e.,

\begin{equation}
\tilde R(s,a) = R(s,a){{\alpha  + \beta (s,a)} \over {\alpha  + y}},
\label{eq26}
\end{equation}
\begin{equation}
\tilde \gamma  = {y \over {\alpha  + y}},
\label{eq27}
\end{equation}
\begin{equation}
\tilde P(s'|s,a) = \left\{ \begin{array}{l}
1 - \frac{{[1 - P(s|s,a)]\beta (s,a)}}{y},s' = s\\
\frac{{P(s'|s,a)\beta (s,a)}}{y},s' \ne s
\end{array} \right.
\label{eq28},
\end{equation}
where $y$ is much larger than $\beta (s,a)$ and is calculated as $y = N{\lambda _p} + {\lambda _v} + {\mu _v} + \sum\nolimits_{i = 1}^N {{f_i}}  + {{M{N_R}{f_v}} \mathord{\left/{\vphantom {{M{N_R}{f_v}} d}} \right.\kern-\nulldelimiterspace} d}$.

Substituting Eqs. \eqref{eq26}-\eqref{eq28} into Bellman equation \cite{D6}, it calculates the normalized maximum value function of each state $s$ for iteration $l+1$ by
\begin{equation}
{\tilde v_{l + 1}}(s) = \mathop {\max }\limits_{a \in \mathbb{A}} \left[ {\tilde R(s,a) + \tilde \gamma \sum\limits_{s' \in \mathbb{S}} {\tilde P(s'|s,a){{\tilde v}_l}(s')} } \right].
\label{eq29}
\end{equation}
Then the absolute difference of this function of each state between two consecutive iterations, i.e.,  $\left\| {{{\tilde v}_{l + 1}}(s) - {{\tilde v}_l}(s)} \right\|$, is calculated. The algorithm is terminated if the absolute difference is smaller than a very small positive threshold, i.e.,
$\left\| {{{\tilde v}_{l + 1}}(s) - {{\tilde v}_l}(s)} \right\| < {{\varepsilon (1 - \tilde \gamma )} \mathord{\left/{\vphantom {{\varepsilon (1 - \tilde \gamma )} 2}} \right.\kern-\nulldelimiterspace} 2}\tilde \gamma$.
Otherwise, the algorithm moves to the next iteration until it is terminated.

When the algorithm is terminated, it outputs the optimal policy under each state, i.e.,
\begin{equation}
{\pi ^*} = \mathop {\arg \max }\limits_{a \in \mathbb{A}} \left[ {\tilde R(s,a) + \tilde \gamma \sum\limits_{s' \in \mathbb{S}} {\tilde P(s'|s,a)\tilde v(s')} } \right].
\label{eq30}
\end{equation}
The optimal offloading strategy is the actions under the optimal policy ${\pi^*}$.

The pseudo-code of the value iteration algorithm is shown in Algorithm 1.

\begin{algorithm}
  \caption{Value Iteration Algorithm}
  \label{al1}
  \KwIn{${s, \mathbb{A},R(s,a), P(s'|s,a), {{\varepsilon (1 - \tilde \gamma )} \mathord{\left/{\vphantom {{\varepsilon (1 - \tilde \gamma )} 2}} \right.\kern-\nulldelimiterspace} 2}\tilde \gamma}$}
  \KwOut{the optimal policy ${\pi ^*}$}
  Initialize $v(s){\rm{ = }}0$, $k = 0$, for each $s \in S$\;
  \For{each $s \in \mathbb{S}$}
  {
    ${\tilde v_{l + 1}}(s) = \mathop {\max }\limits_{a \in \mathbb{A}} \left[ {\tilde R(s,a) + \tilde \gamma \sum\limits_{s' \in S} {\tilde P(s'|s,a){{\tilde v}_l}(s')} } \right]$\;
   }
   \If{$\left\|{{{\tilde v}_{l + 1}}(s)-{{\tilde v}_l}(s)}\right\|<{{\varepsilon (1 - \tilde \gamma )}\mathord{\left/{\vphantom{{\varepsilon (1 - \tilde \gamma )}2}}\right.\kern-\nulldelimiterspace}2}\tilde \gamma$}
    {
       \For{each $s \in \mathbb{S}$}
        {${\pi ^*} = \mathop {\arg \max }\limits_{a \in \mathbb{A}} \left[ {\tilde R(s,a) + \tilde \gamma \sum\limits_{s' \in \mathbb{S}} {\tilde P(s'|s,a)\tilde v(s')} } \right]$\;
         }
     }
  \Else
   {$l = l + 1$\;
    return to Line 2\;
    }

  return the optimal policy ${\pi ^*}$\;
\end{algorithm}



%

\section{Simulation Results}
\textcolor{black}{In this section, we conduct simulation experiments to validate our strategy. We use two different strategies as benchmarks:}

\begin{itemize}
\item[1)] \textcolor{black}{Greedy strategy, which always selects the maximum available resources in the system to offload tasks.}
\item[2)] \textcolor{black}{Equal probability strategy, where it does not consider the event and each action has an equal probability of being selected as specified in (2).}
\end{itemize}
\textcolor{black}{SMDP-based strategy has the polynomial complexity of $O(N^2)$ \cite{SMDPcomplexity}, while the complexities of the greedy strategy and equal probability strategy are $O(N)$ \cite{GAcomplexity}.}
The experiment tool is MATLAB 2019a and the simulation scenario is described in section \ref{system}. Some key parameters are listed in Table \ref{tab2} by referring to \cite{D4}.

\begin{table}\footnotesize
\caption{Parameters of simulation}
\label{tab2}
\vspace*{-0.2cm}
\centering
\begin{tabular}{|c|c|c|c|}
\hline
\textbf{Parameter} &\textbf{Value} &\textbf{Parameter} &\textbf{Value}\\
\hline
$N$ & 4 & $M$ & 4-10 \\
\hline
${\lambda _v}$ & 9 & ${\mu _v}$ & 8 \\
\hline
${f_v}$ & $350 cycles/s$ & ${f_1}$ & $600 cycles/s$ \\
\hline
${f_2}$ & $660 cycles/s$ & ${f_3}$ & $620 cycles/s$ \\
\hline
${f_4}$ & $650 cycles/s$ & $N_R$ & $ 3$ \\
\hline
${W_{min}}$ & 3 & $m$ & 1 \\
\hline
${E_l}$ & $100ms$ & $\eta $ & 5 \\
\hline
$\zeta $ & 28 & $\alpha $ & $0.1$ \\
\hline
${T_{idle}}$ & $20us$ & $\delta $ & $2us$ \\
\hline
$DIFS$ & $50us$ & $SIFS$ & $10us$ \\
\hline
$Header$ & $400bits$ & ${E[P]}$ & $1920bytes$ \\
\hline
$ACK$ & $240bits$ & $ACKtimeout$ & $292bits$ \\
\hline
$\varepsilon $ & $10$ &  &\\
\hline
\end{tabular}
\vspace{-0.6cm}
\end{table}


In the simulation, $Case{\rm{ }}0$, $Case{\rm{ }}1$ and $Case{\rm{ }}2$ represent the decisions to offload the arriving tasks to the platoon, VFC or to discard the tasks, respectively. $A1$, $A2$ and $A3$ represent that the system allocates one, two and three RUs of the VFC for processing, respectively.

\begin{figure}[htbp]
\centering
\includegraphics[width=0.9\linewidth, scale=1.00]{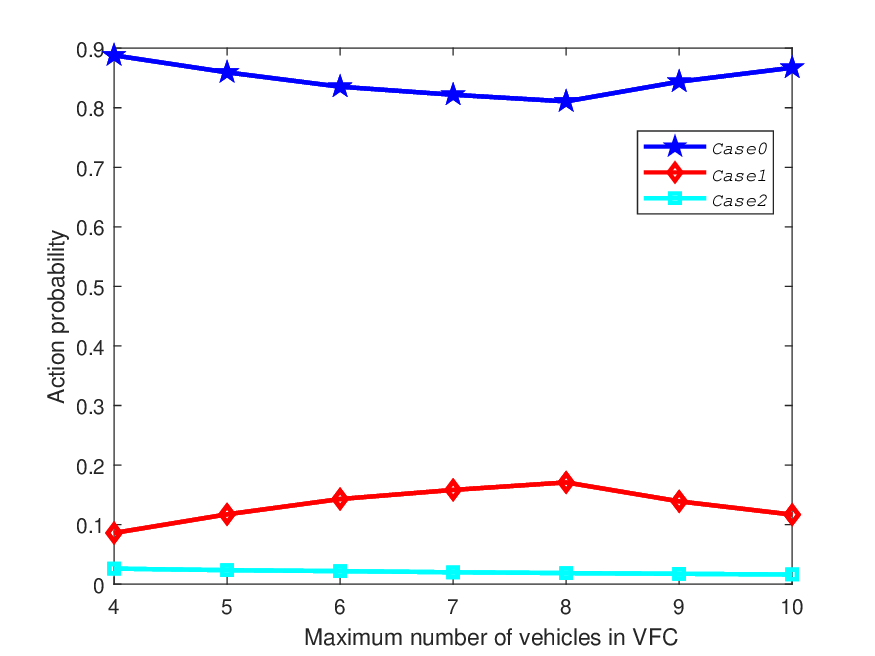}
\caption{\textcolor{black}{Probabilities of $Case{\rm{ }}0$, $Case{\rm{ }}1$ and $Case{\rm{ }}2$ for different maximum number of vehicles in VFC (${\lambda _p}$=20 $task/s$, $d$=40 $cycles$)}}
\label{fig3}
\vspace{-0.2cm}
\end{figure}

Fig. \ref{fig3} shows the probabilities of $Case{\rm{ }}0$, $Case{\rm{ }}1$ and $Case{\rm{ }}2$ for different maximum number of RUs in VFC when the task arrival rate is 20 $task/s$ and $d$ is 40 $cycles$.
We can see that the probability of $Case{\rm{ }}2$ always keeps a small value, which validates that our strategy can process most of tasks with high probability and only discard few tasks. In addition, the probability of $Case{\rm{ }}0$ is always larger than that of $Case{\rm{ }}1$. This is because that the platoon has more available resources than VFC, and the system tends to offload the task to the platoon. Moreover, the probability of $Case{\rm{ }}1$ first gradually increases and the probability of $Case{\rm{ }}0$ first gradually decreases, because the resources in the VFC are gradually increasing with the increase of the maximum number of RUs in the VFC. In this case, more RUs in the VFC can be allocated to process tasks, which decreases the computing delay, thus the system tends to offload the tasks to the VFC to get a larger long-term reward. It also can be seen that the probability of $Case{\rm{ }}0$ is increasing while the probability of $Case{\rm{ }}1$ is decreasing as the maximum number of RUs continues to increase, because as the maximum number of RUs continues to increase, more vehicles in the VFC will face a higher collision probability and thus prolongs the transmitting delay. In this case, the system tends to offload the task to the platoon to get a higher long-term reward.
In summary, there exists one best number of vehicles in VFC to deal with the tasks. This result will help us to control the number of VFC's vehicles when designing VFC in applications.

\begin{figure}[htbp]
\centering
\includegraphics[width=0.9\linewidth, scale=1.00]{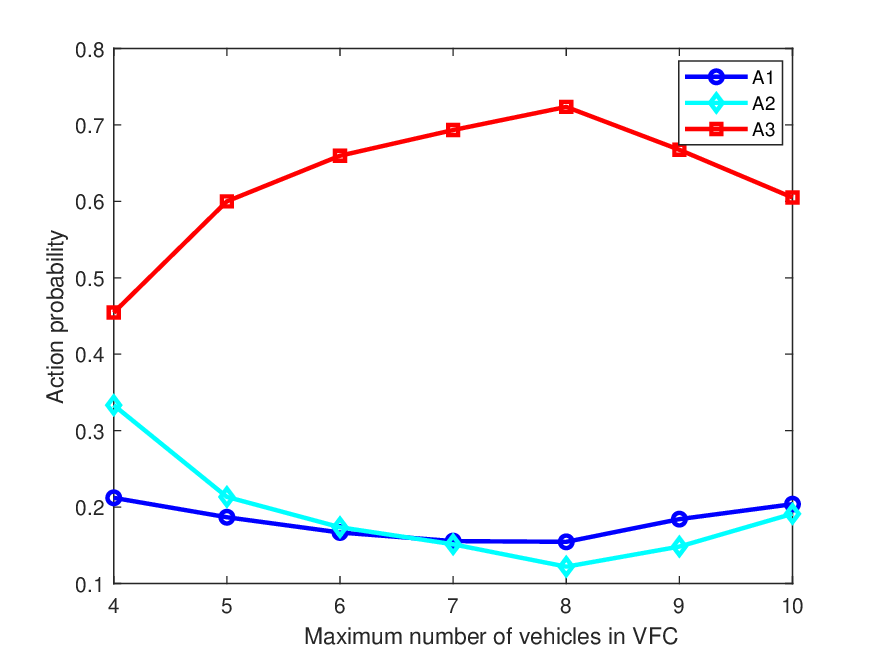}
\caption{Probabilities of  $A1$, $A2$ and $A3$ for different maximum number of \textcolor{black}{vehicles} in VFC (${\lambda _p}$=20 $task/s$, $d$=40 $cycles$)}
\label{fig4}
\vspace{-0.2cm}
\end{figure}
Fig. \ref{fig4} shows the probabilities of $A1$, $A2$ and $A3$ when the task \textcolor{black}{arrival} rate is 20 $task/s$ and $d$ is 40 $cycles$.
We can see that the probability of $A3$ is always larger than those of $A1$ and $A2$, and the probability of $A2$ is larger than that of $A1$ at the beginning. This is because that the VFC needs to process a lot of tasks when the task \textcolor{black}{arrival} rate is 20 $task/s$. The system tends to allocate RUs as many as possible to process the task, and the computing delay can be reduced.
In addition, the probabilities of $A1$ and $A2$ first decrease and the probability of $A3$ first increases with the increase of maximum number of RUs. This is because the resources in VFC gradually increase as the maximum number of RUs increases. The system tends to allocate more RUs in the VFC to reduce the computing delay and obtain a higher long-term reward. Moreover, the probability of $A2$ becomes smaller than that of $A1$ as the maximum number of RUs continues to increase. This is due to that the system allocates more RUs to process tasks due to the high probability of $A3$ in the VFC, which increases the collision probability and transmitting delay. In this case, the system tends to allocate tasks to one RU rather than two RUs to reduce the collision probability.
 We can also observe that the probabilities of $A1$ and $A2$ increase and the probability of $A3$ decreases as the maximum number of RUs continues to increase. This is because that more vehicles transmit data with the increase of the maximum number of RUs,  which increases the collision probability and incurs a large transmitting delay. Thus the system will tend to allocate less RUs to process the task.

\begin{figure}[htbp]
\centering
\includegraphics[width=0.9\linewidth, scale=1.00]{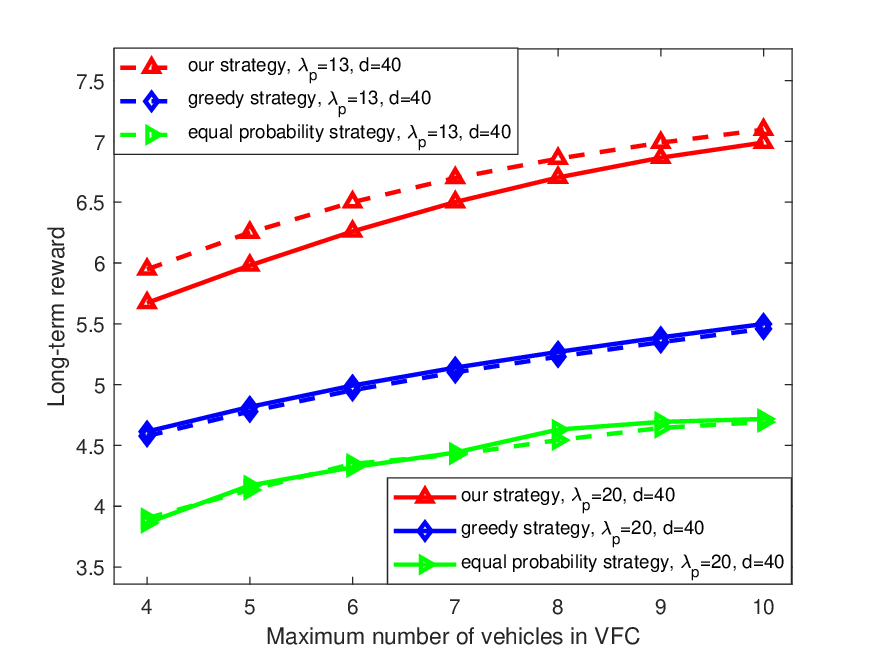}
\caption{\textcolor{black}{Long-term rewards for different maximum number of vehicles in VFC (${\lambda _p} $=13 or 20 $task/s$, $d$=40 $cycles$)}}
\label{fig5}
\vspace{-0.2cm}
\end{figure}

\textcolor{black}{
Fig. \ref{fig5} shows the long-term rewards for different maximum numbers of vehicles in VFC, different strategies and task arrival rates when $d$ is 40 $cycles$. We can see that the long-term reward of the system gradually under different strategies increases with the increase of the maximum number of vehicles in VFC. This is because that the available resource increases as the maximum number of vehicles in VFC increases, which reduces both computing delay and transmitting delay and further improves the long-term reward. In addition, we can find that under different task arrival rates, our proposed strategy can obtain a larger long-term reward than the greedy strategy while the equal probability strategy is inferior to greedy strategy. The reason is that the equal probability strategy does not consider the potential available resources in platoons and VFC, and the greedy strategy always allocates the maximum available resource in the system without considering the long-term reward, while our strategy jointly considers various factors to maximize the long-term reward. Moreover, we can see the long-term reward of our strategy for the task arrival rate $\lambda_p$ being 20 $task/s$ is smaller than that for $\lambda_p$ being 13 $task/s$. This is because that when $\lambda_p$ increases, there are more tasks in the system and thus the time of transmitting and processing tasks becomes longer, which results in smaller long-term reward. We also see that the long-term rewards of the greedy strategy and equal probability strategy are almost the same under different task arrival rates. This is because that the greedy strategy always allocates the maximum available resources for processing, and the equal probability strategy allocates resource based on equal probabilities without considering the variation of the task arrival rate.
}

\begin{figure}[htbp]
\centering
\includegraphics[width=0.9\linewidth, scale=1.00]{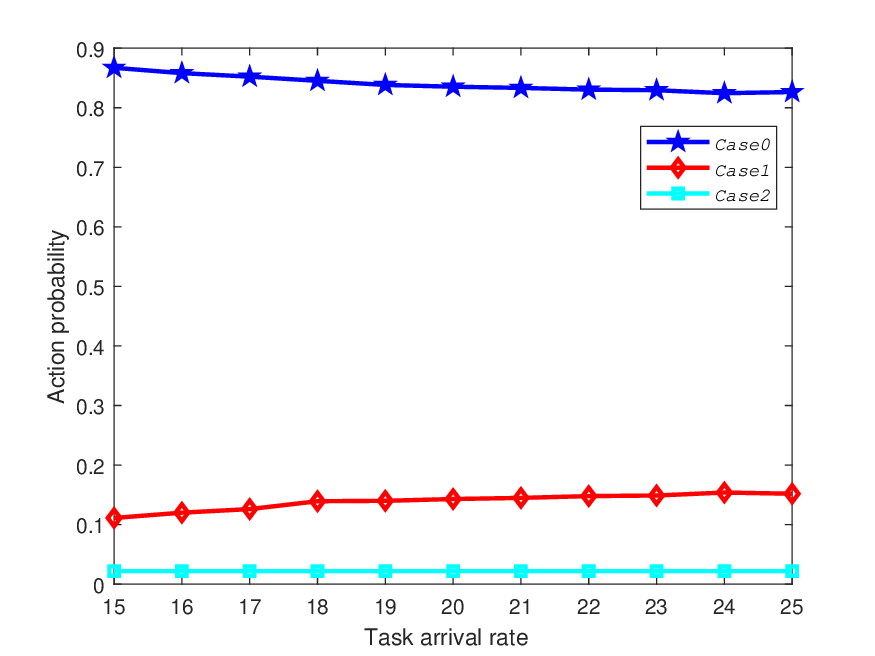}
\caption{\textcolor{black}{Probabilities of $Case{\rm{ }}0$, $Case{\rm{ }}1$ and $Case{\rm{ }}2$ for different task \textcolor{black}{arrival} rates ($d$=40 $cycles$, $K$=6)}}
\label{fig14}
\vspace{-0.2cm}
\end{figure}
Fig. \ref{fig14} shows the probabilities of $Case{\rm{ }}0$, $Case{\rm{ }}1$ and $Case{\rm{ }}2$ for different task \textcolor{black}{arrival} rates when $d$ is 40 $cycles$ and $K$ is 6.
We can see that our system can keep a small probability of discarding the tasks, which validates that our strategy can process most of tasks with high probability and only discard few tasks.
In addition, the probability of $Case0$ is always larger than that of $Case1$. This is because that the platoon has more available resources than VFC, and the system tends to offload the task to the platoon.
When the task \textcolor{black}{arrival} rate is increasing, the probability of $Case0$ is decreasing and the probability of $Case1$ is increasing.
\textcolor{black}{This is because that the resources in the platoon are fixed, and thus when the task arriving rate is large, system tends to allocate more tasks to VFC to process quickly.}

\begin{figure}[htbp]
\centering
\includegraphics[width=0.9\linewidth, scale=1.00]{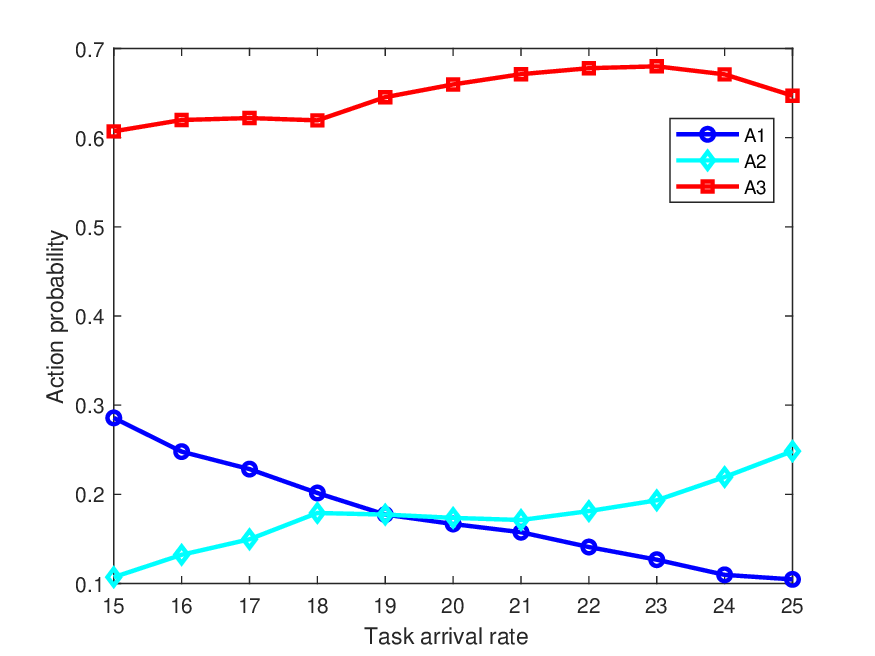}
\caption{Probabilities of  $A1$, $A2$ and $A3$ for different task \textcolor{black}{arrival} rates ($d$=40 $cycles$, $K$=6)}
\label{fig13}
\vspace{-0.2cm}
\end{figure}
Fig. \ref{fig13} shows the probabilities of  $A1$, $A2$ and $A3$ for different task \textcolor{black}{arrival} rates when $d$ is 40 $cycles$ and $K$ is 6. \textcolor{black}{We can see that when the task \textcolor{black}{arrival} rate increases, the probability of  $A1$ decreases, the probability of $A2$ increases, and the probability of  $A3$ first increases and then decreases. This is becase that when the task arrival rate is small, the collision probability is small and allocating more RUs will lead to fast task processing. As the task arrival rate keeps increasing, allocating 3 RUs will lead a large collision probability because that each task will be divided to 3 subtasks to transmit. In this case, the negative impact caused by the collision probability will lead a large delay, which outweights the positive impact of fast task processing. Therefore, the probability of $A3$ decreases when the task arrival rate is large.}

\begin{figure}[htbp]
\centering
\includegraphics[width=0.9\linewidth, scale=1.00]{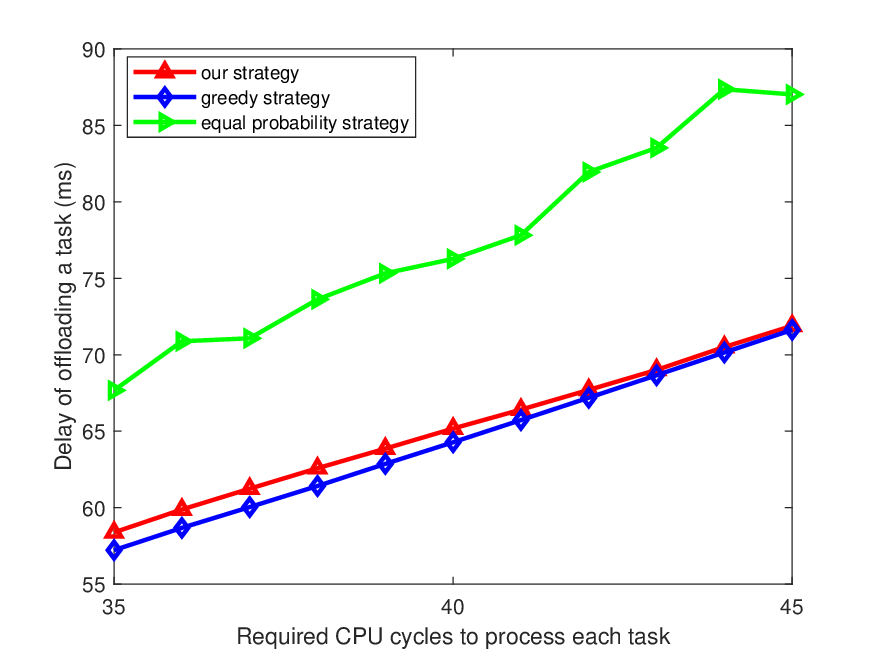}
\caption{Delay of offloading a task for different required CPU cycles to process each task (${\lambda_p}$=20 $task/s$, $K$=6)}
\label{fig17}
\vspace{-0.2cm}
\end{figure}
\textcolor{black}{
Fig. \ref{fig17} shows the delay of offloading a task for different required CPU cycles to process each task and different strategies when $\lambda_p$ is 20 $task/s$ and $K$ is 6. We can see that the offloading delay under different strategies increases with the increasing of required CPU cycles to process each task. This is the time of processing a task becoming longer, thus the offloading delay is also increased. We also see that the delay of the equal probability strategy is the highest and the delay of our strategy is slightly higher than that of the greedy strategy. This is because that the equal probability strategy allocates resource randomly without considering the performance including delay and resources consumption. In addition, our strategy has a slightly higher delay than that of the greedy strategy. This is because the optimization object of our strategy is not only the offloading delay but also the resource occupancy, which is detailed in Eq. \eqref{eq14}. However, the system adopting greedy algorithm always chooses the available resource to minimize the offloading delay without considering the resource occupancy, thus its offloading delay is lower but its long-term reward is inferior to that of our strategy.
}

\begin{figure}[htbp]
\centering
\includegraphics[width=0.9\linewidth, scale=1.00]{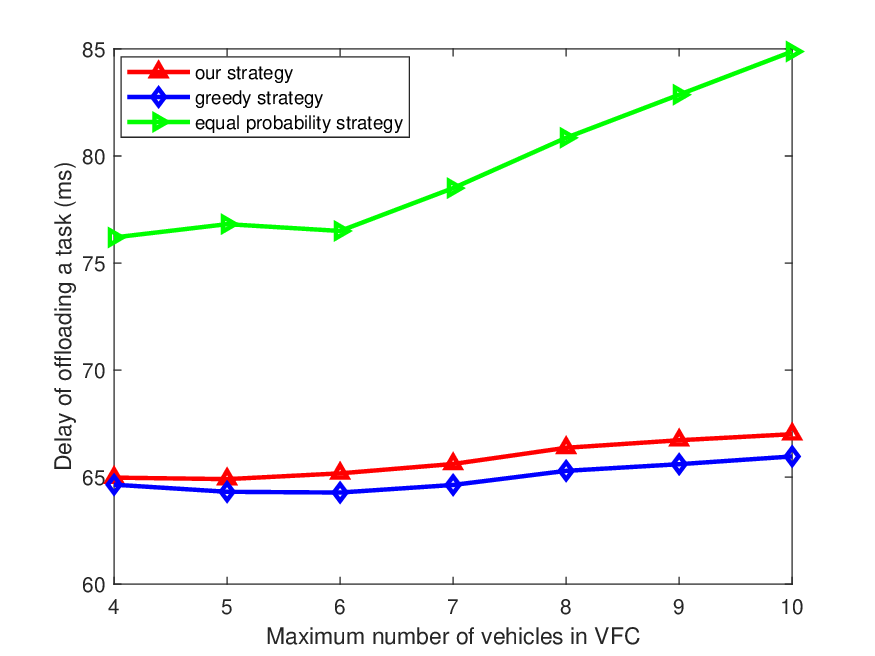}
\caption{Delay of offloading a task for different maximum number of vehicles in VFC (${\lambda_p}$=20 $task/s$, $d$=40)}
\label{figvfcdelay}
\vspace{-0.2cm}
\end{figure}

\textcolor{black}{
Fig. \ref{figvfcdelay} shows the delay of offloading a task for different maximum numbers of vehicles and different strategies in VFC when $\lambda_p$ is 20 $task/s$ and $d$ is 40 $cycles$. We can see that the delay of different strategies increases with the maximum number of vehicles in VFC increases. It is because that with the number of vehicles in VFC increases, the collision probability increases, which causes more transmitting delay. Moreover, the delay of equal probability strategy is the highest and the delay of greedy algorithm is slightly lower than that of our strategy. This is due to the same reason as shown in Fig. \ref{fig17}.
}

\section{Conclusions}
In this paper, we jointly considered the 802.11p DCF mechanism, heterogeneous computation resource of vehicles in the platoon, random task arrival and departure as well as the random arrival and departure of vehicles in VFC. Then we proposed an offloading strategy to obtain the maximal long-term reward based on the SMDP. We first adopted the SMDP to model the offloading process and designed the SMDP framework including state set, action set, reward function and transition probability, where the transmitting delay are derived based on to the 802.11p DCF mechanism to determine the reward. Then we adopt the value iteration algorithm to solve the SMDP model to obtain the optimal offloading strategy. Extensive experiments have been conducted to demonstrate the outperformance of our strategy. According to the theoretical analysis and simulation results, the following conclusions are summarized:
\begin{itemize}
\item Our strategy can process most of tasks and discard few tasks. In addition, the system prefers to offload to the platoon because the available resources in the platoon are larger than that in VFC. Moreover, there are more available RUs as the resources in the VFC increase, thus the system tends to offload task to the VFC. As the resources in the VFC keep increasing, the system tends to offload to the platoon due to the increase of the collision probability in the VFC.

\item The system under our strategy tends to allocate more RUs in the VFC for processing as increase of the resources in the VFC, but as the resources in the VFC keep increasing, the system tends to allocate less RUs for processing due to the increased collision probability in the platoon.


\item Our strategy performs better than the greedy strategy because the 802.11p DCF, heterogeneous computation resource of vehicles in the platoon, the random task arrival and departure as well as random arrival and departure of vehicles in VFC have been jointly considered to get the maximal long-term reward.


\item When the task is allocated to more RUs, the total delay which includes the transmitting delay in the platoon, transmitting delay form the leader vehicle to VFC and the computing delay in the RUs is small.

\item When $d$ and $K$ are fixed, different task \textcolor{black}{arrival} rates will influence the action probability of $A1$, $A2$, $A3$, $Case0$, $Case1$ and $Case2$.
The offloading delay is increasing with the increase of required CPU cycles due to the increasing computing delay in the RUs.

\end{itemize}

\section*{Acknowledgment}
The authors are indebted to Wenhua Wang and Jiahou Chu with Jiangnan University, for their help with this work.

\ifCLASSOPTIONcaptionsoff
  \newpage
\fi

\end{document}